\documentclass[journal=jacsat,manuscript=article]{achemso}

\usepackage[version=3]{mhchem} 

\newcommand{\affD}{SASTRA University, Tirumalaisamudram, Thanjavur, Tamilnadu-613401, India}

\author{Anirban Polley}
\email{anirban.polley@gmail.com}
\affiliation{\affD}

\title[An \textsf{achemso} demo]{Concentration-Dependent Membrane Destabilization in DPPC Bilayers: Distinct Insertion Mechanisms and Stress Redistribution by Chloroform and Alkanols}
\abbreviations{MD, DPPC, CHCl$_3$, MeOH, EtOH, OcOH}

\keywords{lipid bilayers, membrane destabilization, partitioning, lateral stress profile, molecular dynamics, alkanols, chloroform}

\begin{document}


%
%
%


\begin{tocentry}
\centering
\includegraphics[width=8.25cm,height=3.5cm,keepaspectratio]{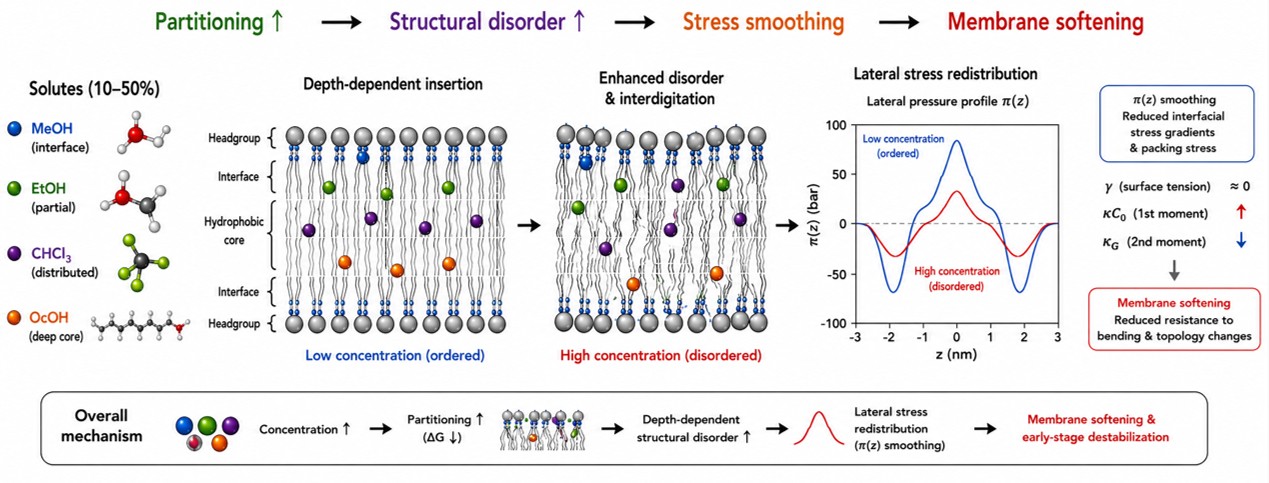}
\end{tocentry}
TOC Caption: Concentration-dependent partitioning of small molecules induces depth-dependent structural disorder and lateral stress redistribution, leading to membrane softening.

\begin{abstract}
How do solute concentration and molecular chemistry govern the transition from membrane saturation to destabilization? We address this using microsecond-scale molecular dynamics simulations of dipalmitoylphosphatidylcholine (DPPC) bilayers with chloroform (CHCl$_3$) and a homologous series of alkanols (methanol, ethanol, octanol) over $0-50\%$ concentrations. Although complete membrane melting is not observed within $1000\, ns$, all systems exhibit clear precursors of destabilization, including enhanced thickness fluctuations, reduced lipid order, and mechanical softening. Chloroform induces pronounced thinning and large fluctuations, consistent with deep, transient insertion. Methanol perturbs primarily the headgroup region, while ethanol shows intermediate behavior with partial insertion. Octanol preserves bilayer thickness at high concentrations due to lipid-like insertion but significantly increases fluctuations and interdigitation. Across all systems, increasing concentration decreases the area compressibility modulus and deuterium order parameter, accompanied by smoothing of lateral pressure profiles, indicating stress redistribution. Free energy analysis reveals increased membrane partitioning and reduced translocation barriers with concentration, strongest for octanol and weakest for methanol. These results demonstrate that membrane destabilization is governed by the interplay of insertion depth, interfacial crowding, and lipid packing disruption.
\end{abstract}

\section{Introduction}

Biological membranes are dynamic, heterogeneous assemblies whose structural integrity and function are highly sensitive to the presence of small molecules \cite{seeman_anesthetic}. Amphiphilic and hydrophobic solutes-including anesthetics, short-chain alcohols, and organic solvents-are known to partition into lipid bilayers and modulate their physical properties \cite{Regen_2009,Ueda_1998,Ueda_2001,anirban_cpl13,anirban_pre18,anirban_pre2025}. Experimental studies have consistently shown that at sufficiently high concentrations, small amphiphilic and hydrophobic molecules destabilize lipid bilayers by increasing permeability, reducing mechanical stability, and perturbing lipid packing, ultimately leading to loss of bilayer integrity \cite{Rowe1983,Chin1977,Cantor_1997a,cantor2001,Evans1987}. Despite extensive investigation, the molecular pathways \cite{Frank_1994,Frank_1997} that connect solute partitioning to membrane destabilization remain incompletely understood.

A central challenge lies in disentangling the relative roles of solute chemistry, insertion depth, and concentration \cite{Rowe1998,Holte1998,Kranenburg2004,Cantor_1997a}. Small molecules such as methanol, ethanol, octanol, and chloroform exhibit markedly different physicochemical characteristics-ranging from hydrophilic to strongly hydrophobic-and therefore interact with membranes through distinct mechanisms. Methanol predominantly localizes near the headgroup region, perturbing interfacial hydrogen-bonding networks \cite{anirban_pre18}, whereas longer-chain alcohols such as octanol can penetrate deeply into the hydrophobic core and adopt lipid-like conformations. Chloroform, in contrast, is highly permeable and exhibits weakly localized, transient insertion within the bilayer interior \cite{anirban_pre18}. While these qualitative trends are well recognized, a quantitative and unified description of how such differences translate into concentration-dependent membrane destabilization is lacking.

From a biophysical perspective, membrane destabilization is not a single event but a progressive process involving changes in lipid packing, thickness, elastic moduli, and lateral stress distribution \cite{mcintosh,Rawicz2000,suryabrahmam2025}. Key observables such as the area compressibility modulus, deuterium order parameter, interleaflet coupling (interdigitation), and lateral pressure profiles provide complementary insights into membrane mechanics. However, most prior studies have focused either on low solute concentrations or on equilibrium structural properties, without systematically probing the high-concentration regime where destabilization becomes prominent. Moreover, the relationship between free energy landscapes of solute insertion and emergent mechanical responses of the membrane remains insufficiently explored.

In this work, we address these gaps using extensive molecular dynamics simulations of dipalmitoylphosphatidylcholine (DPPC) bilayers in the presence of chloroform (CHCl$_3$) and a homologous series of alkanols (methanol, ethanol, and octanol) across a broad concentration range ($0-50\%$). By combining structural, thermodynamic, and mechanical analyses, we characterize how solute concentration and chemistry jointly regulate membrane behavior. Specifically, we quantify changes in membrane thickness and its fluctuations, area compressibility, lipid order parameters, interdigitation, and lateral pressure profiles, alongside free energy barriers and membrane partitioning.

Our results reveal that membrane destabilization is governed not simply by the extent of solute partitioning, but by the coupling between insertion depth, interfacial crowding, and disruption of lipid packing. We demonstrate that distinct classes of solutes follow different pathways--from interfacial perturbation to deep core insertion--leading to qualitatively different modes of mechanical softening and stress redistribution. These findings provide a unified molecular framework for understanding concentration-dependent membrane destabilization and offer broader insight into the action of membrane-active small molecules, including anesthetics and solvents.

\section{\label{sec:level2}Methods}
\noindent
{\it Model membrane}\,:\,

A symmetric, single-component dipalmitoylphosphatidylcholine (DPPC) lipid bilayer is constructed and simulated in an explicit aqueous environment using atomistic molecular dynamics as implemented in GROMACS 5.1 \cite{Lindahl}. The bilayer is equilibrated at a temperature of $50^{\circ} C$ ($323 \,K$), corresponding to the fluid phase of DPPC.

To investigate concentration-dependent effects, chloroform (CHCl$_3$) and a homologous series of alkanols--methanol (MeOH), ethanol (EtOH), and octanol (OcOH)-are introduced into the aqueous phase on both sides of the membrane. The solute concentration relative to lipid content is varied systematically over the range $x=0\%$, $10\%$, $20\%$, $30\%$, $40\%$, and $50\%$.

Each system consists of $256$ lipids ($128$ per leaflet) and $8192$ water molecules, corresponding to a water-to-lipid ratio of 32:1, ensuring full hydration of the bilayer \cite{schwille_dopc_dppc_chol}. Solutes are initially distributed in the aqueous phase and allowed to partition spontaneously into the membrane during the course of the simulations.\\

\noindent
{\it Force fields}\,:\,

Force-field parameters for DPPC are taken from previously validated united-atom lipid models \cite{kindt_dopc_dppc_chol,anirban_pre2025,anirban_jpcb12,anirban_cpl13,Tieleman-POPC,mikko}. Parameters for chloroform and alkanols of varying chain length (methanol, ethanol, and octanol) are adopted from established and widely used models \cite{bockmann_alkanol,anirban_cpl13,MikkoBPJ2006,ramon_jpcb2011,ramon_plosone2013,anirban_pre2025,anirban_pre18}. Water molecules are modeled using the SPC/E model, which incorporates an average polarization correction to the potential energy. 

A united-atom force field is employed to access the long time scales and system sizes required to capture membrane structural fluctuations and lateral heterogeneity. In this representation, nonpolar groups (--CH$_3$, --CH$_2$--) are treated as single interaction sites, while polar functional groups such as --OH are represented explicitly, enabling an accurate description of hydrogen-bonding interactions between alkanols and lipid headgroups \cite{anirban_jcb14,anirban_cell15,kamal2023}.

United-atom lipid force fields are extensively validated for reproducing structural, thermodynamic, and mechanical properties of lipid bilayers, including lateral pressure profiles and membrane elastic moduli. In the present systems, the dominant interactions governing solute behavior include hydrogen bonding at the membrane interface and hydrophobic interactions within the lipid core. These interactions are well captured within the united-atom framework, providing an appropriate balance between computational efficiency and physical accuracy for the phenomena investigated in this study.
\\

\noindent
{\it Initial configurations}\,:\,

Initial configurations of the symmetric multicomponent bilayer systems are generated using PACKMOL \cite{packmol}. Each system consists of a preassembled DPPC bilayer containing $256$ lipids ($128$ per leaflet), hydrated with $8192$ water molecules. Lipids are arranged in a bilayer geometry, while water molecules occupy the surrounding aqueous regions, ensuring full hydration.

Alkanols (methanol, ethanol, and octanol) and chloroform molecules are introduced into the aqueous phase on both sides of the bilayer at varying concentrations. Specifically, $26$, $50$, $76$, $102$, and $128$ solute molecules are added, corresponding to increasing solute-to-lipid ratios. The solutes are initially distributed symmetrically in the aqueous regions and allowed to partition spontaneously into the membrane during equilibration.\\


\noindent
{\it Simulation protocol and equilibration}\,:\,

Each system is initially equilibrated for $50\, ps$ in the NVT ensemble using a Langevin thermostat to remove unfavorable steric contacts. This is followed by simulations in the NPT ensemble at a temperature of $323$ K ($50^{\circ}$C) and a pressure of $1\, atm$.

During the first $100\, ns$ of NPT simulations, the Berendsen thermostat and barostat are employed for equilibration. Subsequently, the Nose-Hoover thermostat and Parrinello-Rahman barostat are used to ensure proper sampling of the isothermal-isobaric ensemble. Pressure coupling is applied semi-isotropically, with a compressibility of $4.5\times 10^{-5}$ bar$^{-1}$.

To improve statistical reliability, four independent simulations are performed for each system, yielding a total sampling time of $4 \,\mu s$ per membrane composition. Each trajectory is propagated for $1000\, ns$, and the final $500\, ns$ are used for analysis.

Long-range electrostatic interactions are treated using the reaction-field method with a cutoff of $r_c=2\, nm$, while Lennard-Jones interactions are truncated at $1\, nm$ \cite{anirban_jpcb12,mikko,patra2004}.

\noindent
{\it Computation of Membrane Elastic Properties}\,:\,

Lateral pressure profiles across the bilayer are computed using the Irving-Kirkwood contour with a spatial resolution of $0.1\, nm$. The pressure tensor is evaluated from pairwise force contributions by rerunning the trajectories with an extended electrostatic cutoff of 2 nm to ensure accurate stress calculations.

All bond lengths are constrained using the LINCS algorithm for lipids and the SETTLE algorithm for water molecules, allowing the use of a $2\,fs$ integration time step \cite{Hess,SETTLE}.

Pressure profiles are accumulated over the final $500\, ns$ of each trajectory. The resulting lateral stress profile is used to compute membrane elastic properties through its moments with respect to the bilayer midplane. In particular, the first moment yields the product $\kappa C_0$ (bending rigidity-spontaneous curvature), while the second moment provides an estimate of the Gaussian curvature modulus $\kappa_G$ \cite{samuli_prl2009}. This mechanical route directly links spatial stress distributions to membrane elasticity and curvature energetics.
\\

\noindent
{\it Statistical Analysis and Error Estimation}\,:\, 

For each system, four independent simulations are performed, each of duration $1000\, ns$, with configurations saved every $100\, ps$. To ensure equilibration, only the final $500\, ns$ of each trajectory ($5000\,$ frames) are used for analysis.

For each replica, observables are first averaged over time to obtain a single representative value. Ensemble-averaged quantities are then computed by averaging these values across the four independent simulations.

Statistical uncertainties are estimated as the standard error of the mean (SEM), defined as 
$SEM=\frac{\sigma}{\sqrt{N}}$, where $\sigma$ is the standard deviation of the observable across the replica averages and $N=4$ is the number of independent simulations.

The distributions of the measured observables are verified to be approximately Gaussian, supporting the use of SEM as a reliable estimate of statistical uncertainty under the assumptions of the Central Limit Theorem. Block averaging analysis is performed to confirm that the chosen sampling window exceeds the autocorrelation time of the observables.

\section{Results and discussion}

\begin{figure*}[h!t]
\begin{center}
\includegraphics[width=16.0cm]{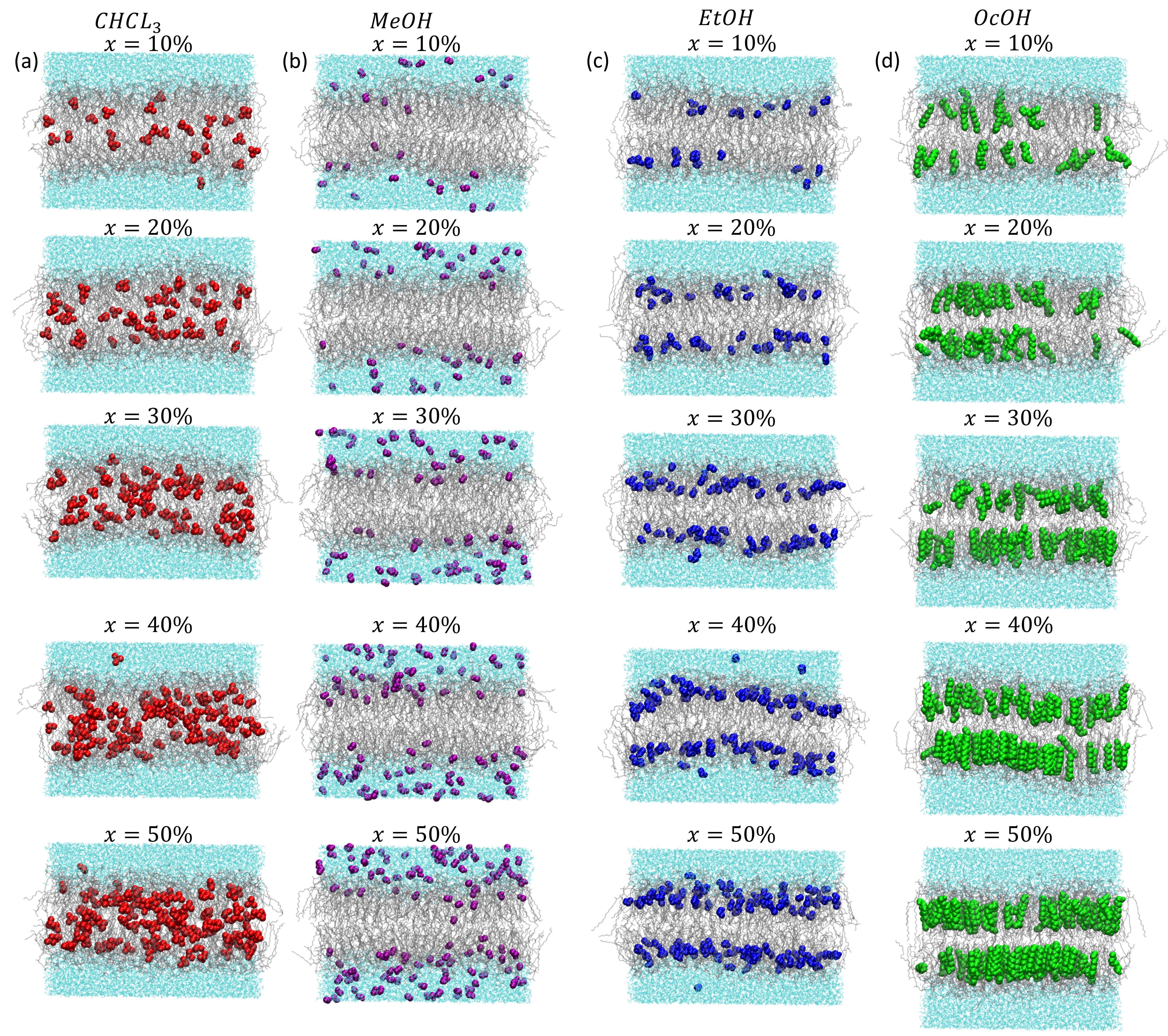}
\caption{Representative equilibrated snapshots ($t=1000\,ns$) of DPPC bilayer membranes containing chloroform and 1-alkanols (methanol, ethanol, and octanol) at concentrations $x=10\%$, $20\%$, $30\%$, $40\%$ and $50\%$, shown for each system.
}
\label{fig1}
\end{center}
\end{figure*}

We perform four independent simulations for each system, yielding a total sampling time of $4\,\mu s$ per membrane composition. Each trajectory is propagated for $1000\,ns$, and the final $500\,ns$ are used for analysis. Representative snapshots of DPPC bilayers without anesthetic molecules (as shown in Fig.~S1 in the Supplementary Information (SI)) and with chloroform and alkanols (methanol, ethanol, and octanol) at varying concentrations ($x=10\%$, $20\%$, $30\%$, $40\%$, and $50\%$) are shown in Fig.~\ref{fig1} and Fig.~S2 in SI, respectively.

Equilibration of each system is assessed through the time evolution of the area per lipid and the total energy, as presented in Fig. S3 in SI. Both observables reach stable plateaus prior to the production window, confirming adequate equilibration \cite{anirban_jcb14,anirban_jpcb12,anirban_cpl13}.

\subsection{Structural Instability of the Membrane}

{\it Membrane Thickness and Fluctuations:}

The membrane thickness is computed as the average distance between the phosphate headgroup planes of the two leaflets. Specifically, the bilayer is divided into lateral grids, and the local thickness $d(x,y)$ is defined as the distance between the centers of mass of the headgroup atoms (phosphorus atoms) in the upper and lower leaflets \cite{anirban_pre2025,anirban_jcp2026arxiv}. The mean membrane thickness (shown in Fig. ~S4 in SI) is then obtained as

\begin{equation}
    \langle d\rangle=\langle d(x,y)\rangle
\end{equation}

where the averaging is performed over the membrane surface and over time.

To quantify the effect of solutes, we define the relative change in thickness as,
\begin{equation}
    \delta d=\frac{\langle d^{no\, anesthetic}\rangle-\langle d^{anesthetic}\rangle}{\langle d^{no\, anesthetic}\rangle}
\end{equation}

which provides a normalized measure of thinning (positive $\delta d$) or thickening (negative $\delta d$) relative to the pure membrane (shown in Fig. ~S5 in SI).

Thickness fluctuations are characterized by the standard deviation of the local thickness distribution,

\begin{equation}
    \langle \sigma_t\rangle=\sqrt{\langle (d(x,y)-\langle d \rangle)^2\rangle}
\end{equation}

which captures spatial and temporal undulations of the bilayer. An increase in $\sigma_d$  reflects enhanced membrane roughness and serves as an early indicator of structural instability.

\begin{figure*}[h!t]
\begin{center}
\includegraphics[width=16.0cm]{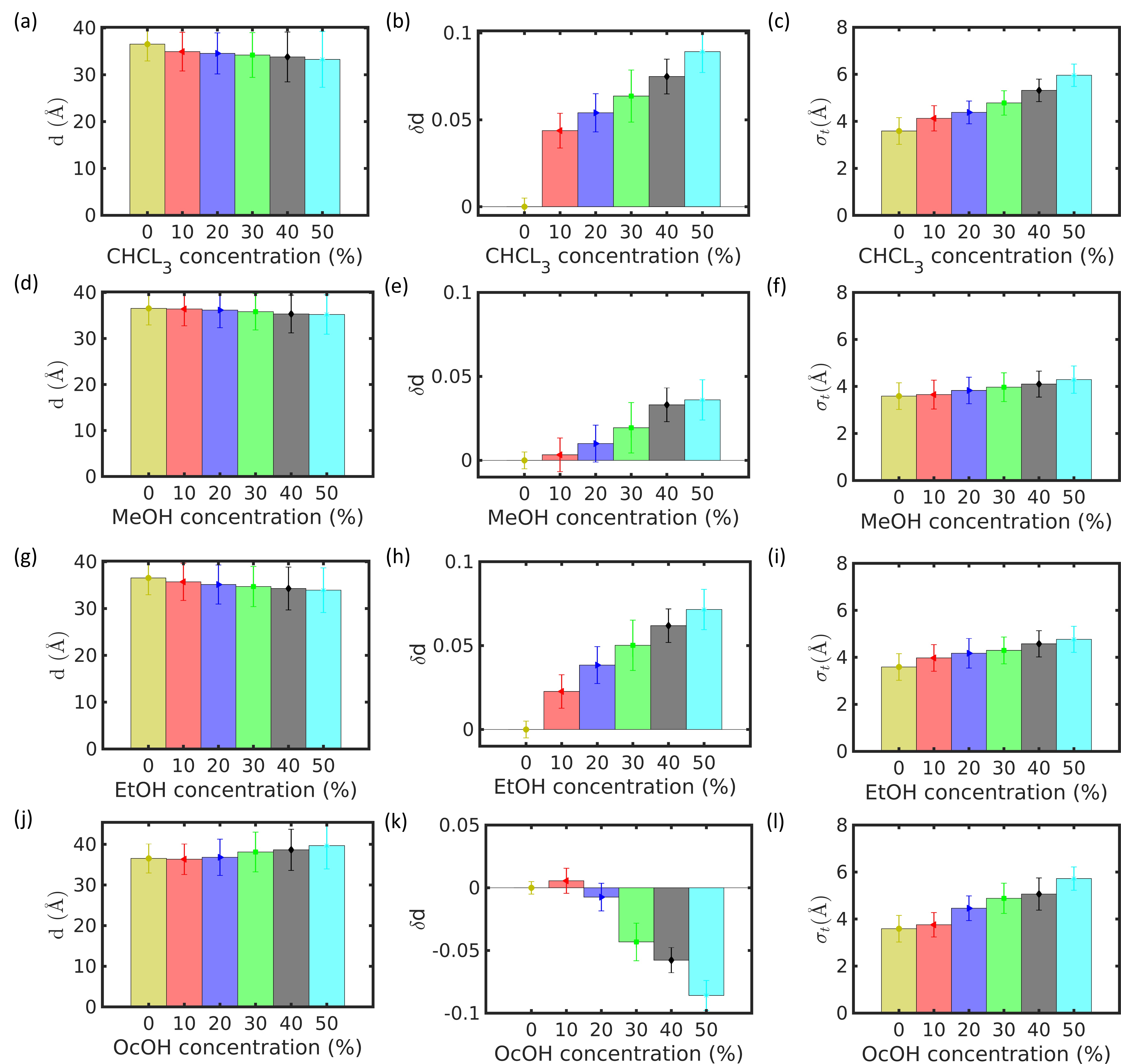}
\caption{Membrane thickness (d), relative change in thickness $\delta d=\frac{\langle d^{no\, anesthetic}\rangle-\langle d^{anesthetic}\rangle}{\langle d^{no\, anesthetic}\rangle}$, and thickness fluctuations $\langle \sigma_t\rangle=\sqrt{\langle (d(x,y)-\langle d \rangle)^2\rangle}$ of DPPC bilayers containing chloroform (CHCl$_3$) and 1-alkanols (methanol, ethanol, and octanol) at concentrations $x=10\%$, $20\%$, $30\%$, $40\%$, and $50\%$. Increasing CHCl$_3$ concentration induces pronounced membrane thinning accompanied by a significant increase in thickness fluctuations. Methanol produces slight thinning with a modest increase in $\sigma_t$, consistent with interfacial perturbations. Ethanol results in moderate thinning and fluctuations, reflecting partial insertion into the bilayer. In contrast, octanol exhibits minimal thinning at low-to-intermediate concentrations and slight thickening at higher concentrations, indicating thickness preservation due to lipid-like insertion, while fluctuations increase across all concentrations.
}
\label{fig2}
\end{center}
\end{figure*}

With increasing concentration of chloroform (CHCl$_3$), the membrane exhibits pronounced thinning (large positive $\delta d$) accompanied by a significant increase in $\sigma_t$, indicating strong disruption due to deep and transient penetration into the bilayer core. In contrast, methanol (MeOH) induces only slight thinning (small $\delta d$) and a marginal increase in $\sigma_t$, consistent with its preferential localization at the headgroup-neck region and limited penetration into the hydrophobic core as shown in Fig.~\ref{fig1}.

Ethanol (EtOH) shows intermediate behavior, producing moderate thinning and a corresponding increase in fluctuations, reflecting partial insertion into the bilayer. Octanol (OcOH), however, displays a non-monotonic response: an initial thinning at low concentration ($10\%$), followed by minimal change at intermediate concentrations ($20\%$), and slight thickening at higher concentrations (negative $\delta d$). This behavior is consistent with lipid-like insertion and hydrophobic matching. Notably, despite this apparent preservation of average thickness, $\sigma_t$ increases with concentration, indicating enhanced fluctuations and underlying structural destabilization.

These results demonstrate that while $\langle d \rangle$ and $\delta d$ capture average structural changes, $\sigma_t$ provides a more sensitive metric for detecting early-stage membrane instability (Fig.~\ref{fig2}).

{\it Area per Lipid and Compressibility:}

\begin{figure*}[h!t]
\begin{center}
\includegraphics[width=16.0cm]{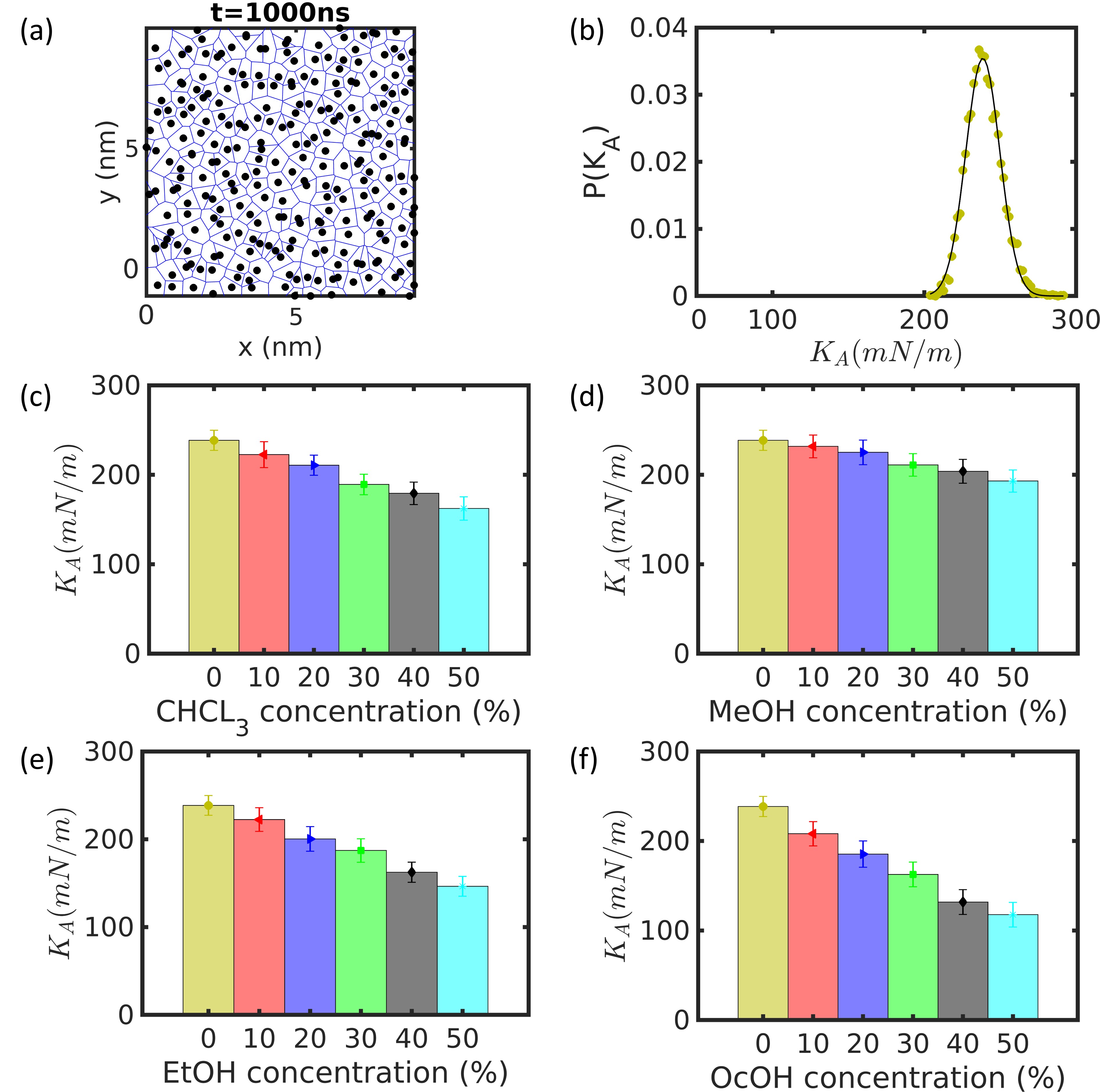}
\caption{(a) Voronoi tessellation used to compute the area per lipid, A(t). The area compressibility modulus is calculated as $K_A=\frac{k_B T <A>}{<(\delta A)^2}$, where $\delta A=A(t)-\langle A \rangle$. (b) Probability distribution of $K_A$ for DPPC bilayers. (c–f) Concentration-dependent variation of $K_A$ for DPPC bilayers containing chloroform (CHCl$_3$) and 1-alkanols (methanol, ethanol, and octanol) at concentrations $x=10\%$, $20\%$, $30\%$, $40\%$, and $50\%$. The time evolution of $K_A$ is shown in the Supplementary Information (SI). Increasing solute concentration leads to a systematic decrease in $K_A$, indicating enhanced area fluctuations and progressive mechanical softening of the membrane.
}
\label{fig3}
\end{center}
\end{figure*}

The area per lipid is computed using Voronoi tessellation \cite{anirban_jpcb12,anirban_cell15,anirban_jcb14,anirban_prl16,anirban_cell15} in the membrane plane as shown in Fig.~\ref{fig3}, where each lipid is assigned a polygonal area based on its lateral coordinates. The instantaneous area per lipid, $A(t)$, is obtained for each frame, and the mean area is given by $\langle A \rangle$.

The area compressibility modulus, $K_A$, is calculated from area fluctuations as

\begin{equation}
K_A=\frac{k_B T \langle A \rangle}{\langle (\delta A)^2 \rangle}
\end{equation}

with $\delta A=A(t)-\langle A \rangle$, where $k_B$ is the Boltzmann constant and $T$ is the temperature (shown in Fig.~S6 in SI). This relation connects thermal fluctuations to membrane elasticity, with larger fluctuations $\langle (\delta A)^2 \rangle$ corresponding to lower compressibility and hence a mechanically softer membrane.

The time evolution of $K_A$ is presented in the Supplementary Information (SI), confirming convergence of the calculated values. The probability distributions of $K_A$ are shown in Fig.~\ref{fig2}(b), while the concentration-dependent variation is summarized in Fig.~\ref{fig3}(c–f).

With increasing concentration of chloroform (CHCl$_3$), methanol (MeOH), ethanol (EtOH), and octanol (OcOH), the area compressibility modulus decreases systematically. This reduction in $K_A$ reflects enhanced area fluctuations and indicates progressive mechanical softening of the membrane. The effect is more pronounced for solutes that penetrate deeper into the bilayer (e.g., octanol and chloroform), whereas interfacially localized solutes such as methanol produce comparatively weaker changes as shown in Fig.~\ref{fig3}.

These results, together with the increase in thickness fluctuations, demonstrate that solute incorporation leads to a reduction in lateral cohesion and an increase in membrane deformability, providing a key signature of early-stage destabilization.

\subsection{Lipid Order and Packing}

{\it Deuterium Order Parameter ($S_{cd}$):}

The orientational order of lipid acyl chains is quantified using the deuterium order parameter, $S_{cd}$, defined as

\begin{equation}
    S_{cd}=\frac{1}{2}\langle 3 cos^2 \theta-1 \rangle
\end{equation}

where $\theta$ is the angle between the vector connecting consecutive united-atom sites along the lipid tail and the bilayer normal. The angular brackets denote an ensemble and time average. This definition provides an effective measure of chain ordering consistent with united-atom representations. This definition yields values comparable to experimentally measured deuterium order parameters up to a scaling factor.

To assess the effect of solutes, we define the relative change in the order parameter as

\begin{equation}
    \delta S_{cd}=\frac{S_{cd}^{anesthetic}-S_{cd}^{no\,anesthetic}}{S_{cd}^{no\,anesthetic}}
\end{equation}

which provides a normalized measure of chain disordering upon solute incorporation.

\begin{figure*}[h!t]
\begin{center}
\includegraphics[width=14.0cm]{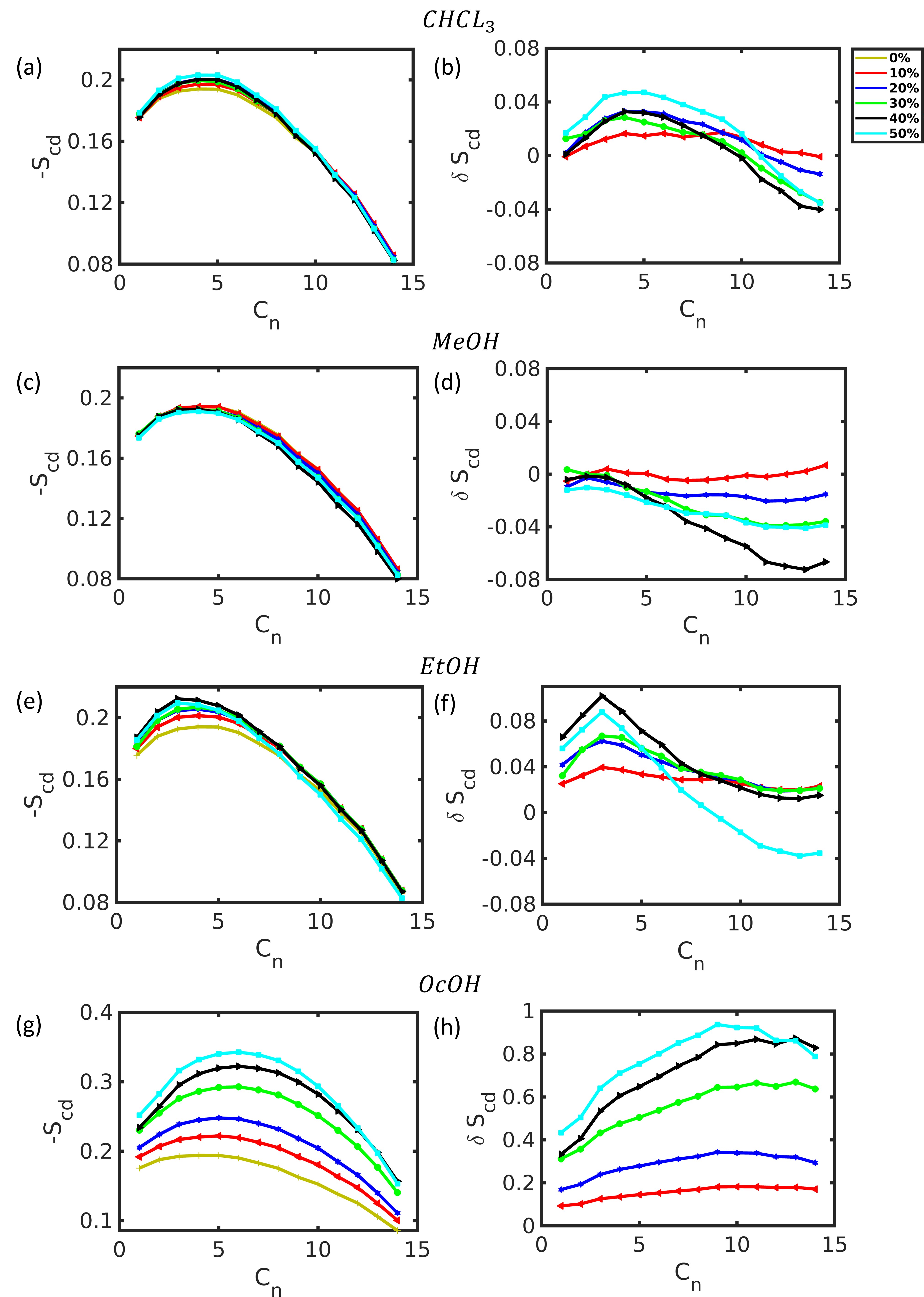}
\caption{
Deuterium order parameter, $S_{cd}$, and the relative change in order parameter, $\delta S_{cd} = \frac{S_{cd}^{\text{anesthetic}} - S_{cd}^{\text{no anesthetic}}}{S_{cd}^{\text{no anesthetic}}}$, for DPPC bilayers in the presence of chloroform (CHCl$_3$) and 1-alkanols (methanol, ethanol, and octanol) at varying concentrations. Increasing solute concentration leads to a systematic decrease in $S_{cd}$ and negative values of $\delta S_{cd}$, indicating progressive disordering of lipid acyl chains.
}
\label{fig4}
\end{center}
\end{figure*}

With increasing concentration of chloroform (CHCl$_3$), methanol (MeOH), ethanol (EtOH), and octanol (OcOH), a systematic decrease in $S_{cd}$ is observed, corresponding to negative values of $\delta S_{cd}$. This reduction indicates progressive disordering of lipid acyl chains and disruption of their orientational alignment along the bilayer normal as shown in Fig.~\ref{fig4}.

The extent of disordering depends on solute chemistry and insertion depth. Methanol, which primarily localizes at the headgroup region, induces only a modest decrease in $S_{cd}$. Ethanol exhibits an intermediate effect due to partial insertion into the bilayer. In contrast, chloroform and octanol produce a more pronounced reduction in $S_{cd}$, consistent with their deeper penetration into the hydrophobic core and stronger perturbation of lipid packing.

The observed decrease in chain order correlates with the reduction in area compressibility and the increase in thickness fluctuations, indicating that lipid disordering is a key molecular origin of membrane softening and destabilization.

{\it Tail Interdigitation:}

Interleaflet coupling is quantified using two complementary measures based on density overlap and chain penetration (shown in Fig.~S7).

The normalized interdigitation, $I_{norm}$, is defined as

\begin{equation}
    I_{norm}=\frac{\int \rho_{up} (z) \rho_{low}(z) dz}{\sqrt{\int\rho_{up}(z)^2 dz}\sqrt{\int\rho_{low}(z)^2 dz}}
\end{equation}

where $\rho_{up}(z)$ and $\rho_{low}(z)$ are the mass density profiles of lipid tails in the upper and lower leaflets, respectively. This dimensionless metric quantifies the spatial overlap and shape similarity of the two density distributions and is sensitive to both leaflet broadening and density redistribution.

To capture direct chain penetration, we also compute the crossing interdigitation,
\begin{equation}
    I_{cross}=\frac{N_{up \rightarrow down}+N_{down \rightarrow up}}{N_{total}}
\end{equation}

where $N_{up \rightarrow down}$ and $N_{down \rightarrow up}$ denote the number of tail atoms crossing the bilayer midplane, and $N_{total}$ is the total number of tail atoms (shown in Fig.~S8 in SI). This metric directly measures interleaflet penetration events.

To quantify solute-induced changes, we define the relative variations

\begin{equation}
    \delta I_{norm}=\frac{I_{norm}^{anesthetic}-I_{norm}^{no\,anesthetic}}{I_{norm}^{no\, anesthetic}},  
    \delta I_{cross}=\frac{I_{cross}^{anesthetic}-I_{cross}^{no\,anesthetic}}{I_{cross}^{no\, anesthetic}} 
\end{equation}

\begin{figure*}[h!t]
\begin{center}
\includegraphics[width=10.0cm]{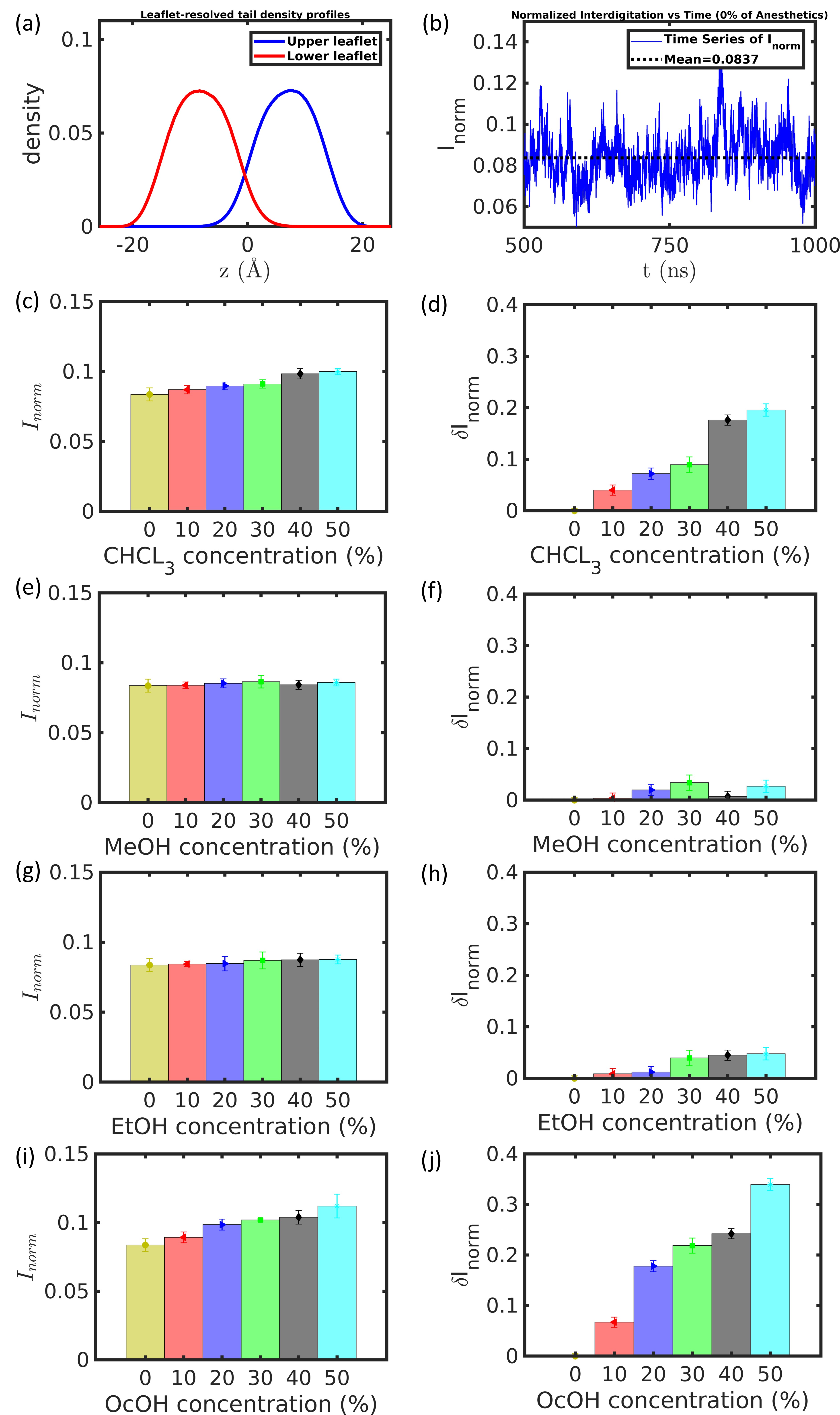}
\caption{
(a) Leaflet-resolved tail density profiles of DPPC lipids. (b) Time evolution of the normalized interdigitation, $I_{\text{norm}}$. (c–h) Concentration-dependent variation of $I_{\text{norm}}$ and the relative change $\delta I_{\text{norm}} = \frac{I_{\text{norm}}^{\text{anesthetic}} - I_{\text{norm}}^{\text{no anesthetic}}}{I_{\text{norm}}^{\text{no anesthetic}}}$ for DPPC bilayers containing chloroform (CHCl$_3$) and 1-alkanols (methanol, ethanol, and octanol). 
With increasing concentration of CHCl$_3$, $I_{\text{norm}}$ increases significantly, accompanied by a corresponding rise in $\delta I_{\text{norm}}$, indicating enhanced interleaflet coupling. Methanol (MeOH) shows minimal variation in $I_{\text{norm}}$ with a slight increase in $\delta I_{\text{norm}}$, consistent with its interfacial localization. Ethanol (EtOH) exhibits a moderate increase in both $I_{\text{norm}}$ and $\delta I_{\text{norm}}$, reflecting partial insertion into the bilayer. Octanol (OcOH) produces a pronounced increase in interdigitation, with significant enhancement in both $I_{\text{norm}}$ and $\delta I_{\text{norm}}$, indicating strong interleaflet penetration.
}
\label{fig5}
\end{center}
\end{figure*}

With increasing concentration of chloroform (CHCl$_3$), both $I_{norm}$ and $I_{cross}$ increase significantly, indicating strong interleaflet coupling driven by disruption of the bilayer core. In contrast, methanol (MeOH) shows minimal change in $I_{norm}$ and only a slight increase in $I_{cross}$, consistent with its interfacial localization and weak perturbation of the hydrophobic region as shown in Fig.~\ref{fig5}.

Notably, $I_{norm}$ captures changes in density overlap and leaflet broadening, whereas $I_{cross}$ directly probes chain penetration across the bilayer midplane. The consistent increase in both measures, particularly for chloroform and octanol, demonstrates that interdigitation is a key structural signature of membrane destabilization. These trends correlate with the reduction in lipid order and enhanced thickness fluctuations, linking interleaflet coupling to membrane softening. The increase in $I_{cross}$ precedes significant changes in average thickness, indicating that interdigitation acts as an early marker of bilayer destabilization.

\subsection{Partitioning and Free Energy Landscape}

{\it Partitioning ($f_{mem}$) and Free Energy Landscape ($\Delta G$):}

The membrane partitioning of solutes is quantified by the fraction of molecules residing within the bilayer,

\begin{equation}
f_{mem}=\frac{N_{mem}}{N_{total}},
\end{equation}

where $N_{mem}$ is the number of solute molecules within the membrane region and $N_{total}$ is the total number of solute molecules.

To obtain a thermodynamic measure of membrane affinity, we define the partition coefficient as

\begin{equation}
K=\frac{\rho_{mem}}{\rho_{water}}
=\frac{N_{mem}/V_{mem}}{N_{water}/V_{water}},
\end{equation}

where $\rho_{mem}$ and $\rho_{water}$ are the number densities of solute molecules in the membrane and aqueous regions, respectively.

The corresponding free energy of partitioning is given by

\begin{equation}
\Delta G_{part}=-k_B T \ln K,
\end{equation}

which quantifies the thermodynamic preference of the solute for the membrane relative to the aqueous phase.

To characterize insertion and translocation across the bilayer, we further define the free energy barrier,

\begin{equation}
\Delta G_{barrier}=G(z_{max})-G(z_{min}),
\end{equation}

where $G(z)$ is the free energy profile along the bilayer normal, $z_{min}$ corresponds to the equilibrium free energy minimum within the membrane, and $z_{max}$ denotes the maximum along the translocation pathway.

These quantities are obtained from equilibrium density distributions and free energy profiles along the membrane normal. While $f_{mem}$ provides a direct measure of membrane occupancy, the partition coefficient $K$ and $\Delta G_{part}$ account for differences in accessible volume and permit quantitative comparison across systems (Fig.~S9 in the SI).

\begin{figure*}[h!t]
\begin{center}
\includegraphics[width=10.0cm]{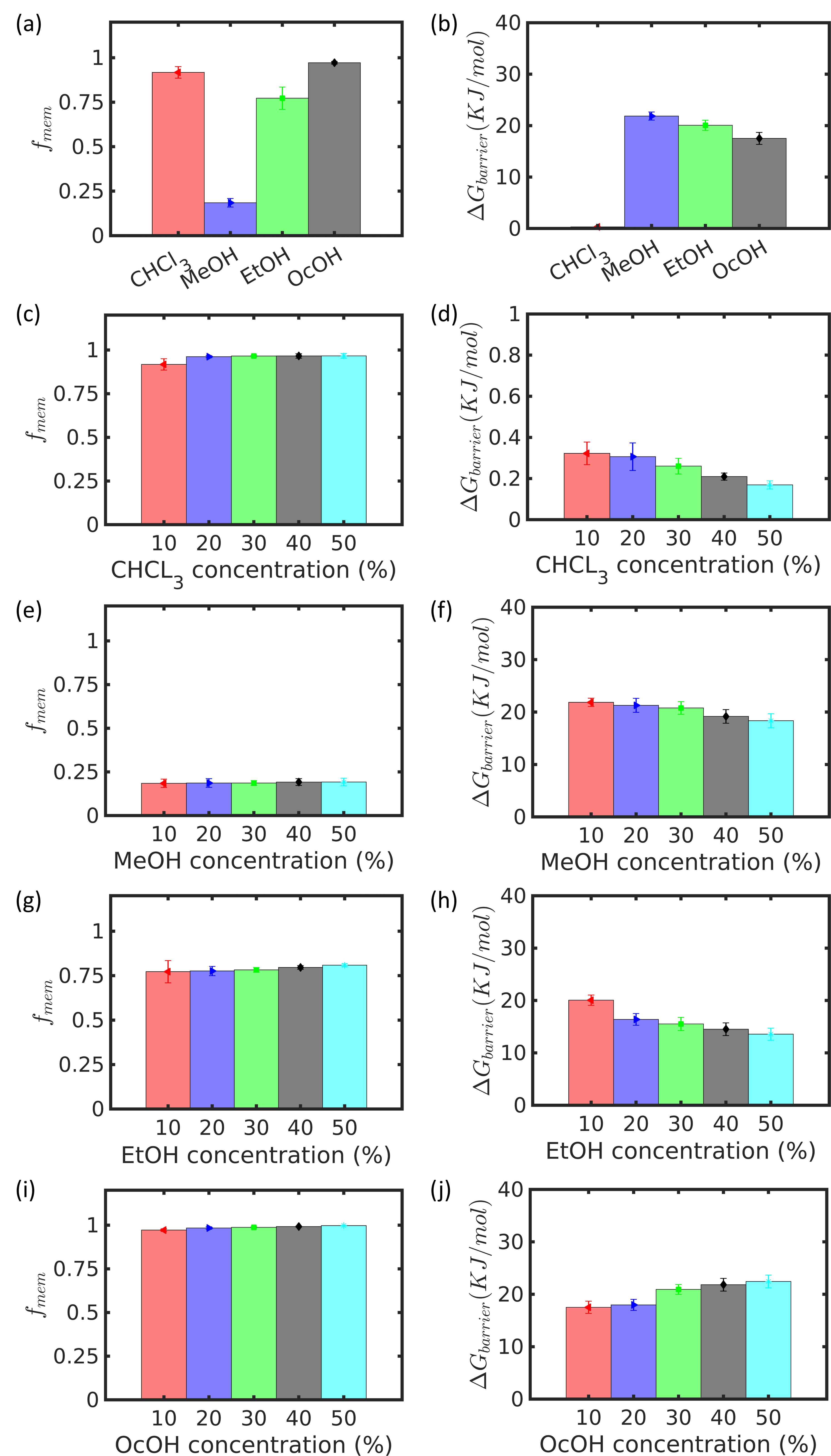}
\caption{
(a,b) Partitioning fraction, $f_{\text{mem}}$, and free energy barrier, $\Delta G_{\text{barrier}}$, for chloroform (CHCl$_3$), methanol (MeOH), ethanol (EtOH), and octanol (OcOH) at $x = 10\%$ in DPPC membranes. (c,e,g,i) Concentration dependence of $f_{\text{mem}}$, and (d,f,h,j) corresponding $\Delta G_{\text{barrier}}$ for each solute. 
At fixed concentration, $f_{\text{mem}}$ decreases in the order OcOH $>$ CHCl$_3$ $>$ EtOH $>$ MeOH, while $\Delta G_{\text{barrier}}$ follows the opposite trend. Increasing solute concentration ($x = 10$–$50\%$) leads to a systematic increase in $f_{\text{mem}}$ and a decrease in $\Delta G_{\text{barrier}}$, indicating enhanced membrane partitioning and reduced insertion barriers. The effect is strongest for octanol, moderate for ethanol, and weakest for methanol, whereas chloroform exhibits weak concentration dependence due to its intrinsically low barrier and high permeability.
}
\label{fig6}
\end{center}
\end{figure*}

A clear dependence on solute chemistry is observed. The partitioning follows the trend \\
OcOH$>$CHCl$_3$$>$EtOH$>$MeOH,
indicating increasing membrane affinity with hydrophobicity. In contrast, the insertion free energy barrier, $\Delta G_{barrier}$, exhibits the opposite ordering, \\
MeOH$>$EtOH$>$OcOH$>$CHCl$_3$, 
reflecting the reduced energetic cost of insertion for more hydrophobic solutes as shown in Fig.~\ref{fig6}.

Increasing solute concentration ($10–50\%$) leads to a systematic increase in $f_{mem}$ and a corresponding decrease in $\Delta G_{barrier}$ for all systems. This behavior indicates enhanced membrane saturation and facilitated translocation at higher concentrations. The magnitude of this effect depends strongly on solute chemistry: it is most pronounced for octanol, moderate for ethanol, and minimal for methanol. Chloroform exhibits relatively weak concentration dependence due to its intrinsically low insertion barrier and high permeability.

The reduction in $\Delta G_{barrier}$ is accompanied by a broadening of the free energy minimum within the bilayer, indicating increased configurational freedom of solutes at higher concentrations.
 
{\it Mechanistic differentiation:}

The observed trends reveal distinct insertion mechanisms governed by solute chemistry.

{\it Octanol (OcOH)}: Strong hydrophobic interactions drive deep insertion into the bilayer core, leading to high partitioning and a significant reduction in $\Delta G_{barrier}$. Its lipid-like alignment stabilizes the membrane core while simultaneously enhancing interleaflet coupling and thickness fluctuations.

{\it Ethanol (EtOH)}: Amphiphilic character promotes accumulation near the membrane interface, resulting in moderate partitioning and a gradual reduction in $\Delta G_{barrier}$. Membrane perturbation arises primarily from interfacial crowding and partial insertion into the hydrophobic region.

{\it Methanol (MeOH)}: Hydrophilic character limits penetration into the bilayer, leading to weak partitioning and comparatively high $\Delta G_{barrier}$. Insertion occurs predominantly through defect-mediated pathways, resulting in weak coupling to the membrane core.

{\it Chloroform (CHCl$_3$)}: Exhibits high permeability with an intrinsically low $\Delta G_{barrier}$, enabling rapid and transient penetration across the bilayer. Its effect is spatially diffuse, producing moderate partitioning but relatively weak concentration dependence.

These results demonstrate that membrane destabilization is governed not only by the extent of partitioning, but also by the coupling between insertion depth, insertion energetics, and lipid packing disruption. Hydrophobic solutes reduce $\Delta G_{barrier}$ and enhance membrane partitioning; however, their impact on membrane stability depends critically on how they redistribute within the bilayer and perturb local membrane organization.

\subsection{Mechanical Stability and Stress Redistribution}

{\it Lateral Pressure Profiles:}

The lateral pressure difference across the membrane \cite{helfrich1973} is defined as

\begin{equation}
\pi(z)=P_L(z)-P_{zz}(z)
\label{lat_press}
\end{equation}

where $P_L(z)=\frac{P_{xx}(z)+P_{yy}(z)}{2}$
is the lateral (in-plane) pressure component and $P_{zz}(z)$ is the normal pressure component. The coordinate z denotes the bilayer normal.

\begin{figure*}[h!t]
\begin{center}
\includegraphics[width=16.0cm]{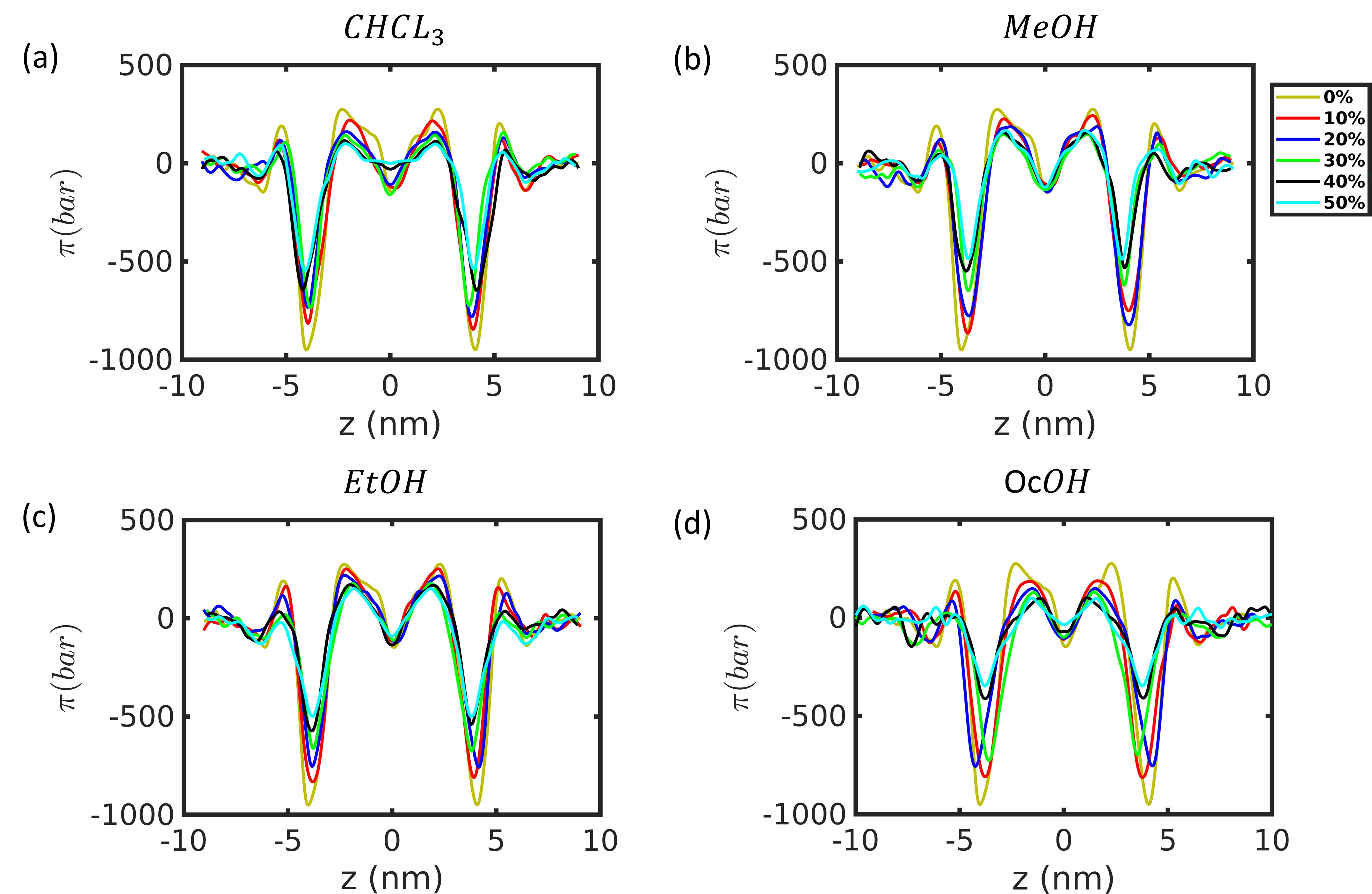}
\caption{
Lateral stress profiles, $\pi(z)$, of DPPC bilayers in the presence of (a) chloroform (CHCl$_3$), (b) methanol (MeOH), (c) ethanol (EtOH), and (d) octanol (OcOH) at concentrations $x = 0\%, 10\%, 20\%, 30\%, 40\%,$ and $50\%$. 
Increasing CHCl$_3$ concentration leads to a moderate reduction in headgroup peaks, a less pronounced negative interfacial stress minimum, and slight changes in the core region, resulting in overall profile smoothing. Methanol produces similar but weaker changes, with minimal perturbation of the core region. Ethanol induces a more noticeable reduction in headgroup stress and a shallower interfacial minimum, accompanied by slight flattening in the bilayer core. Octanol exhibits the strongest effect, with significant reduction in headgroup peaks, substantial weakening of the interfacial stress minimum, and marked smoothing of the core region.
Overall, increasing solute concentration progressively smoothens the lateral stress profile, indicating reduced interfacial stress gradients and lipid packing frustration.
}
\label{fig7}
\end{center}
\end{figure*}

The lateral pressure profile provides spatially resolved information on stress distribution across the membrane and is highly sensitive to solute insertion and lipid packing as shown in Fig.~\ref{fig7}.

With increasing concentration of chloroform (CHCl$_3$), the pressure profile exhibits a moderate reduction in the headgroup peaks, a less pronounced negative dip in the interfacial region, and a slight decrease in the core region, resulting in an overall smoothing of the profile. This behavior reflects its weakly localized and dynamically distributed insertion within the bilayer \cite{Cantor_1997a,cantor2001,franks2008}.

Methanol (MeOH) produces a decrease in the headgroup region and a slight reduction in the interfacial dip, while the core region remains largely unchanged, indicating minimal redistribution of stress and preservation of the overall profile shape. This is consistent with its preferential localization at the membrane interface.

Ethanol (EtOH) induces a more pronounced reduction in the headgroup peaks and a shallower interfacial minimum, accompanied by slight flattening in the core region. These changes suggest partial insertion and inward redistribution of lateral stress.

Octanol (OcOH) exhibits the most significant effect, with a strong reduction in the headgroup peaks, substantial weakening of the interfacial stress minimum, and a marked decrease in the core region. This leads to a highly smoothed pressure profile, consistent with deep insertion and strong perturbation of lipid packing.

Overall, increasing solute concentration leads to progressive smoothing of the lateral pressure profile, reflecting a reduction in interfacial stress gradients and lipid packing stress. The magnitude of this effect follows the insertion depth of the solutes, with octanol and chloroform producing the strongest redistribution, ethanol showing intermediate behavior, and methanol exhibiting the weakest impact.

{\it Surface Tension and Elastic Response:}

Assuming planar symmetry, the surface tension of the bilayer is defined as

\begin{equation}
\gamma=\int_{z_1}^{z_2} \pi(z)\,dz.
\label{gamma}
\end{equation}

where $\pi(z)=P_L(z)-P_{zz}(z)$
is the lateral stress profile, and the integration spans the entire bilayer thickness from 
$z_1$ to $z_2$. 

Membrane elastic properties are related to the moments of the lateral stress profile. For a monolayer, these are given by

\begin{equation}
\kappa C_0 = \int_{z_m}^{z_w} (z-\delta)\pi(z)\,dz,
\label{KC}
\end{equation}

\begin{equation}
\kappa_G = \int_{z_m}^{z_w} (z-\delta)^2 \pi(z)\,dz,
\label{KG}
\end{equation}

where $z_m$ is the bilayer midplane and $z_w$ corresponds to the bulk solvent region where $\pi(z)\rightarrow0$. The integration is performed over a single leaflet, from the bilayer midplane to the bulk water interface. The parameter $\delta$ denotes the neutral plane of the monolayer. For symmetric bilayers, both leaflets contribute equally, and the reported values correspond to bilayer-averaged quantities.

{\it Stress Redistribution and Mechanical Stability:}

Alcohols induce a progressive redistribution of lateral stress from the headgroup and interfacial regions toward a more uniform profile across the bilayer. This effect is strongest for octanol (OcOH), consistent with its deep penetration into the hydrophobic core. Ethanol (EtOH) produces moderate redistribution due to partial insertion, whereas methanol (MeOH) primarily perturbs the headgroup region with minimal impact on the core. Chloroform (CHCl$_3$) induces moderate but spatially diffuse changes, reflecting its weakly localized and highly permeable insertion.

Increasing solute concentration leads to systematic smoothing of the lateral stress profile, indicating a reduction in interfacial stress gradients and lipid packing frustration. The magnitude of this effect follows the insertion depth of the solutes, with octanol showing the strongest impact, followed by ethanol and chloroform, and methanol exhibiting the weakest effect.

The surface tension remains close to zero for all systems, consistent with simulations performed in a tensionless ensemble. For low solute concentrations, $\gamma$ fluctuates around zero within statistical uncertainty, while a small positive deviation is observed at higher concentrations, indicating slight imbalance in lateral stress as shown Fig.~S10 in SI.

{\it Elastic Moduli ($\kappa C_0$ and $\kappa_G$) Trends:}

\begin{figure*}[h!t]
\begin{center}
\includegraphics[width=10.0cm]{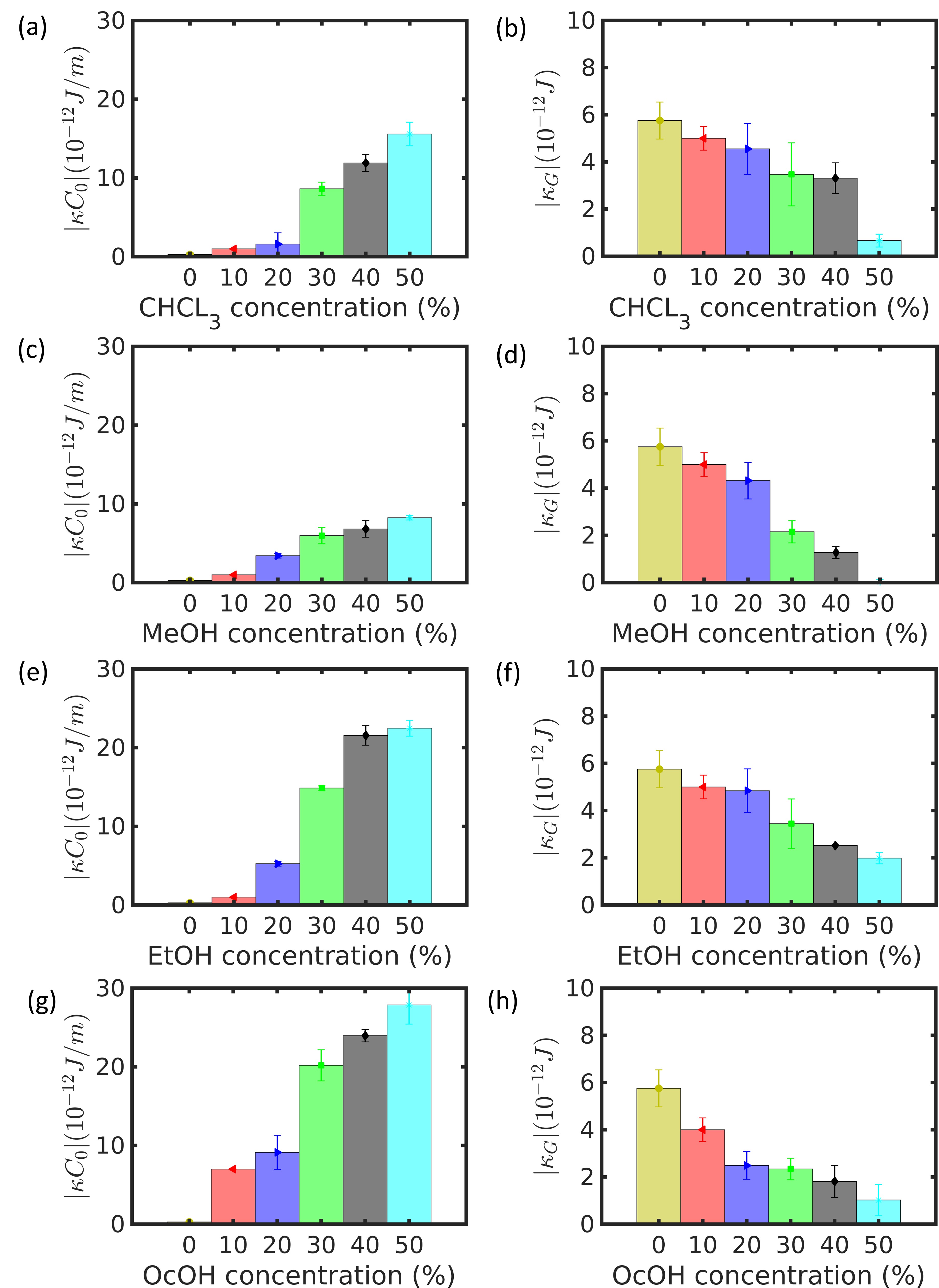}
\caption{
First moment ($\kappa C_0$) and second moment ($\kappa_G$) of the lateral stress profile for DPPC bilayers in the presence of chloroform (CHCl$_3$), methanol (MeOH), ethanol (EtOH), and octanol (OcOH) at concentrations $x = 0\%, 10\%, 20\%, 30\%, 40\%,$ and $50\%$. Panels (a,c,e,g) correspond to $\kappa C_0$, and (b,d,f,h) correspond to $\kappa_G$ for CHCl$_3$, MeOH, EtOH, and OcOH, respectively.
The first moment, $\kappa C_0$, increases systematically with solute concentration for all systems, with the magnitude following the trend OcOH $>$ EtOH $>$ CHCl$_3$ $>$ MeOH, indicating enhanced curvature stress with increasing insertion depth. In contrast, the second moment, $\kappa_G$, decreases with concentration for all solutes, reflecting a reduced resistance to topological deformations and increased membrane softness.
}
\label{fig8}
\end{center}
\end{figure*}

The first moment, $\kappa C_0$, increases with solute concentration for all systems, indicating enhanced curvature stress and increasing asymmetry in the lateral stress distribution. The magnitude follows the trend OcOH$>$EtOH$>$CHCl$_3$$>$MeOH, consistent with increasing insertion depth and perturbation of the bilayer interior as shown in Fig.~\ref{fig8}. 

In contrast, the second moment, $\kappa_G$, decreases systematically with concentration, indicating a reduced resistance to topological deformations such as pore formation or membrane remodeling as shown in Fig.~\ref{fig8}.

The observed decrease in $\kappa_G$, together with increased interdigitation, suggests a reduced energetic barrier for defect formation, consistent with early-stage membrane destabilization.

\section{Conclusion}

We establish a unified, concentration-dependent mechanism that links solute partitioning, structural perturbation, and mechanical response in DPPC membranes \cite{matsumoto_bba2024}. Increasing solute concentration enhances membrane occupancy ($f_{\text{mem}}$) and increases the partition coefficient ($K$), thereby lowering the insertion free-energy barrier ($\Delta G_{\text{barrier}}$) for membrane penetration and translocation. The magnitude and concentration dependence of these effects are governed by solute chemistry, following the hydrophobicity-driven trend:
OcOH $>$ CHCl$_3$ $>$ EtOH $>$ MeOH.

Enhanced partitioning directly translates into structural disruption of the membrane. Deeply inserting solutes such as octanol and chloroform induce significant chain disorder (decrease in $S_{cd}$), enhanced thickness fluctuations ($\sigma_d$), and increased interleaflet coupling (increase in $I_{\text{norm}}$ and $I_{\text{cross}}$). Ethanol produces intermediate effects through partial insertion and interfacial crowding, whereas methanol primarily perturbs the headgroup region with comparatively weak disruption of the bilayer core.

These structural perturbations manifest as mechanical softening of the membrane. Increasing solute concentration progressively smoothens the lateral stress profile, $\pi(z)$, indicating reduced interfacial stress gradients and lipid packing frustration. This redistribution of stress is accompanied by an increase in the first moment ($\kappa C_0$), reflecting enhanced curvature stress, and a decrease in the second moment ($\kappa_G$), indicating reduced resistance to topological deformations. The surface tension remains close to zero across all systems, consistent with the tensionless ensemble employed in the simulations.

Taken together, these results define a coherent pathway for membrane destabilization in which concentration-driven partitioning lowers insertion barriers, promotes depth-dependent structural disorder and interdigitation, redistributes lateral stress, and ultimately softens the membrane. Importantly, membrane destabilization is governed not solely by the extent of membrane partitioning, but by the coupling between insertion depth, lipid packing disruption, and stress redistribution. These findings provide a unified physical framework for understanding how small molecules modulate membrane stability and membrane-mediated biophysical processes.

\begin{acknowledgement}

A.P. appreciates the hospitality of generous computing facilities at  SASTRA University, Thanjavur, Tamilnadu. 
A.P. acknowledges the support under Science and Engineering Research Board (SERB), Department of Science and Technology, Government of India [SERB-SRG/2022/001489] and T.R. Rajagopalan research fund, SASTRA University, India.

\end{acknowledgement}

\begin{suppinfo}

Supplementary Information (SI) contains $10$ figures, including initial configurations of the membrane and time evolution of key properties such as area per lipid, total energy, area compressibility, partitioning, surface tension, and crossing interdigitation.

\end{suppinfo}


\providecommand{\latin}[1]{#1}
\makeatletter
\providecommand{\doi}
  {\begingroup\let\do\@makeother\dospecials
  \catcode`\{=1 \catcode`\}=2 \doi@aux}
\providecommand{\doi@aux}[1]{\endgroup\texttt{#1}}
\makeatother
\providecommand*\mcitethebibliography{\thebibliography}
\csname @ifundefined\endcsname{endmcitethebibliography}
  {\let\endmcitethebibliography\endthebibliography}{}

\end{document}